# Quantum protocols for transference of proof of zero-knowledge systems

José Cláudio do Nascimento and Rubens Viana Ramos

*Abstract* — Zero-knowledge proof system is an important protocol that can be used as a basic block for construction of other more complex cryptographic protocols. An intrinsic characteristic of a zero-knowledge systems is the assumption that is impossible for the verifier to show to a third part that he has interacted with the prover. However, it has been shown that using quantum correlations the impossibility of transferring proofs can be successfully attacked. In this work we show two new protocols for proof transference, being the first one based on teleportation and the second one without using entangled states.

## I. INTRODUCTION

A zero-knowledge system is a protocol between two partners where one of them is the prover, P, and the other is the Verifier, V. Basically, P has a secret (for example, he knows how to solve an specific problem) and V wants to be sure about that. Thus, V challenges P (asking him to solve the problem) and P has to take actions that proof to V that he really has the secret without allowing V to discover it. Besides the zero-knowledge characteristic, the protocol requires an additional property known as *impossibility of transferring the proof*. This property states that, at the end of the protocol, V can not prove to a third part that he had interacted with P. The most famous zero-knowledge proof system is the Goldreich, Micali and Wigderson (GMW) systems [1]. It can be shortly described as follows: P and V know two graphs $G_0$ and $G_1$ with $n$ nodes. P has as secret the isomorphism $\sigma$: $G_1 \rightarrow G_0$. In order to P proofing to V that he really knows $\sigma$, the following protocol is used:

1. P generates a random isomorphism $\lambda$:$G_0 \rightarrow H$ and he sends $H$ to V.
2. V generates a random bit $b$ and sends it to P.
3. P sends the isomorphism $\xi = \lambda \circ \sigma^b$ to V.
4. V checks if $\xi(G_b) = H$. ($\lambda \circ \sigma^b G_b$ : if $b=0 \Rightarrow \lambda G_0 = H$ and if $b=1 \Rightarrow \lambda \circ \sigma G_1 = \lambda G_0 = H$)

The steps are repeated $n$ times and V will only believe in P if step 4 is always satisfied. The probability of P to deceive V with success decreases fast when $n$ grows.

José Cláudio do Nascimento and Rubens Viana Ramos are with Department of Teleinformatic Engineering, Federal University of Ceara, Fortaleza, Ceará, Brazil, e-mails: claudio@deti.ufc.br, rubens@deti.ufc.br. This work was supported by the Brazilian agency FUNCAP.

The impossibility of transferring the proof of the GMW system is known to be resistant against classical attacks, however, it was shown in [2] that it is not resistant against quantum attacks, according to the following protocol: Firstly, a third part, hereafter known as Eve, is colluded with V. They share in advance $n^2$ pairs of the quantum state $(|00\rangle_{VE}+|11\rangle_{VE})/2^{1/2}$ and they agree in a Hash function $h$ that maps all existents graphs with $n$ nodes in a $n$ bit sequence. The steps of the protocol are:

1. P generates a random isomorphism $\lambda$:$G_0 \rightarrow H$ and he sends $H$ to V.
2. V calculates $h(H)$ obtaining a $n$ bit sequence $B=\{B_1,B_2,\ldots,B_n\}$. V then takes $n$ qubits (from the $n^2$ pairs that he shares with Eve) and performs measurements in them. For the $i$-th qubit ($i=1,\ldots,n$) if $B_i$ is '0' then the measurement is performed in the rectangular basis $\{|0\rangle,|1\rangle\}$ otherwise, the basis used is the diagonal $\{(|0\rangle+|1\rangle)/2^{1/2},(|0\rangle-|1\rangle)/2^{1/2}\}$. The states $|0\rangle$ and $(|0\rangle+|1\rangle)/2^{1/2}$ code bit '0' while the states $|1\rangle$ and $(|0\rangle-|1\rangle)/2^{1/2}$ code bit '1'. Thus, the results of the measurements form another $n$-bit sequence, $R=\{R_1,R_2,\ldots,R_n\}$. At last, the bit $b$ that V sends to P is the parity of $R$, $R_1 \oplus R_2 \oplus R_3 \oplus \ldots \oplus R_n$.
3. P sends the isomorphism $\xi = \lambda \circ \sigma^b$ to V.
4. V checks if $\xi(G_b) = H$.
5. Steps 1-4 are repeated $n$ times and it is supposed that V really verifies that P has the secret. After that, V sends to Eve all graphs $H$ (in the right order $H_1,\ldots,H_n$) and all isomorphisms $\xi$ (in the right order $\xi_1,\ldots,\xi_n$) received from P, and all bits $b$ (in the right order $b_1,\ldots,b_n$) sent to P.
6. For the $k$-th $H$ received, $H_k$, $k=1,2,\ldots,n$, Eve calculates $h(H_k)$ obtaining a $n$ bit sequence $B'_k = \{B'_{1k}, B'_{2k}, \ldots, B'_{nk}\}$. Now, Eve takes the $k$-th set of $n$ qubits (from the same pairs used by V when P sent $H_k$ to him) and she performs measurements in them. For the $i$-th measurement, if $B'_{ik} = '0'$ then the rectangular basis is used otherwise, the diagonal basis is used. The results of Eve's measurements form another $n$-bit sequence, $R'$. Eve then calculates the parity of $R'$, $b'$, and checks if $\xi_k(G_{b'}) = H_k$. If this last condition is satisfied for all $n$ $H$ received, then Eve believes that V has interacted with P.

For each time that V challenges P $n$ states $(|00\rangle_{VE}+|11\rangle_{VE})/2^{1/2}$ are used, because the function Hash maps the graphs in a $n$-bit sequence. If V challenges P $n$ times, then $n^2$ states are required. If all random variables used are uniformly distributed, then the probability of V to deceive Eve is $1/2^n$.

The security of the just described protocol is strongly based on the difficulty of inverting the Hash function. However, it has been shown that Hash functions can be efficiently inverted if a quantum computer is used [3,4]. In fact, it seems that the protocol proposed in [2] that permits V to transfer the proof to Eve is not secure against a quantum attack. More precisely, if V uses a quantum computer, he can deceive Eve, making her to believe that he has interacted with P without having done it. Let us suppose that V did not interact with P but he wants to make Eve believe that he has interacted. For this, we assume that V knows all collisions of the agreed Hash function. In order to V deceiving Eve, he must provide to her $n$ tuples $(H,\xi,b)_1^n$ with an appropriated probability distribution of the components, and the components $H_i$, $\xi_i$ and $b_i$ must satisfy $\xi_i G_{b_i} = H_i$ for all $i=1,\ldots,n$. In order to get this, V follows the following steps:

1. V chooses randomly, with uniform distribution, a bit $c$.
2. V chooses, with uniform distribution, a random isomorphism $\xi:G_c \to H$.
3. V calculates $h(H)$ and uses its bits to choose the bases of the measurements, whose results form the bit sequence $R$.
4. V calculates the parity of $R$ and compares to $c$. If they are equal, what happens with probability ½, V sends $(H, \xi, c)$ to Eve. If they are not equal, V takes a collision $H'$, $h(H')=h(H)$, calculates $\xi'$ such that $\xi' G_{\bar{c}} = H'$ and sends $(H', \xi', \bar{c})$ to Eve.

As can be seen from step 4, in order to have success in cheating Eve, V must know $n$ pairs of collisions $(H, H')$ of the Hash function $h$ and $n$ pairs of isomorphism $(\xi, \xi')$ such that $\xi:G_0 \to H$ and $\xi':G_1 \to H'$. Now, the important question is: what is the difficult of V finding the collisions of $h$? If V is supposed to have only classical computers, then the task will be really hard and one can say that the protocol proposed in [2] is secure. On the other hand, if V has a quantum computer he can find the collisions in an efficient way. It has been shown that using a quantum computer with Grover's quantum search algorithm, one is capable of find a collision with an upper bound of $O(n^{1/3})$ [3] and a lower bound of $O(n^{1/5})$ [4] on the number of queries needed. Thus, if V has a quantum computer able to run Grover's search algorithm and he knows $n$ pairs of isomorphism $(\xi, \xi')$ such that $\xi:G_0 \to H$ and $\xi':G_1 \to H'$ where $H$ and $H'$ are collision of $h$, then V can cheat Eve. Hence, the loophole of the protocol proposed in [2] is the use of the Hash function. In order to circumvent this problem, one can use a permutation instead of a Hash function. Since the permutation preserves the number of bits, there will not be collisions and the protocol becomes secure. However, the price to be paid for this security is a larger number of necessary bipartite states. If there are $k$ graphs with $n$ nodes, and the protocol is repeated $m$ times, then the number of bipartite states to be used is $m2^k$.

## II. A NEW QUANTUM ATTACK FOR TRANSFERENCE OF PROOF OF ZERO-KNOWLEDGE SYSTEMS

In this section it is described a different quantum attack, based on teleportation, for the transference of proof of zero-knowledge systems. Firstly, let us consider a classically controllable quantum operator that act as $U|\theta\rangle=|\varphi\rangle$ where $|\theta\rangle$ is one of the single-qubits states $|0\rangle$, $|1\rangle$, $|+\rangle$ and $|-\rangle$ ($|\pm\rangle=(|0\rangle\pm|1\rangle)/2^{1/2}$). Thus, the operator $U$ for a particular bit sequence $C$, $U_c$, can be described by the set of pairs input-output $S=\{(|0\rangle,U_c|0\rangle),(|1\rangle,U_c|1\rangle),(|+\rangle,U_c|+\rangle),(|-\rangle,U_c|-\rangle)\}$. A collision happens when two different bit sequences leads to the same set $S$. In order to realize the new quantum attack, it is assumed that V and Eve share in advance $n$ pairs of the Bell state $(|00\rangle_{VE}+|11\rangle_{VE})/2^{1/2}$ and that Eve has sent to V $n$ qubit states belonging to the set $\{|0\rangle,|1\rangle, |+\rangle,|-\rangle\}$. The new quantum attack for the transference of proofs of zero-knowledge systems is as follows:

1. P generates a random isomorphism $\lambda:G_0 \to H$ and he sends $H$ to V.
2. V calculates $h(H)$ obtaining a $n$ bit sequence $C=\{C_1,C_2,\ldots,C_n\}$. V then takes one qubit $|\theta\rangle$ (from the $n$ qubits that Eve sent to him) and performs a unitary transformation in it according to the bit sequence $C$. The output state $U_c|\theta\rangle=|\varphi\rangle$ is teleported to Eve using one of the $n$ Bell states that Eve and V share. At last, the bit $b$ that V sends to P is the xor function between the bits $D_0$ and $D_1$ obtained in the Bell measurement during teleportation protocol.
3. P sends the isomorphism $\xi=\lambda \circ \sigma^b$ to V.
4. V checks if $\xi(G_b)=H$.
5. Steps 1-4 are repeated $n$ times and it is supposed that V really verifies that P has the secret. After that, V sends to Eve all graphs $H$ (in the right order $H_1,\ldots,H_n$) and all isomorphisms $\xi$ (in the right order $\xi_1,\ldots,\xi_n$) received from P.
6. For the $k$-th $H$ received, $H_k$, $k=1,2,\ldots,n$, Eve calculates $h(H_k)$ obtaining a $n$ bit sequence

$C'_k = \{C'_{1k}, C'_{2k}, \ldots, C'_{nk}\}$. Now, Eve applies the inverse of $U_{c'}$ in the teleported qubit that she received from V in order to recover the $k$-th single-qubit that she sent to V. Once Eve knows exactly the qubit value that she sent, she can perform a measurement in the correct basis and she knows what result is expected to happen. At last, Eve uses the $b$ value that she received during teleportation stage (the xor function between the two classical bits of the teleportation) and checks if $\xi_k(G_b)=H_k$. If this last condition is satisfied for all $n$ $H$ received, then Eve believes that V has interacted with P.

Now, let us suppose that V did not interact with P but he wants to make Eve believe that he has interacted. For this, once more we assume that V knows all collisions of the agreed Hash function $h$. In order to V deceiving Eve, he must provide to her $n$ tuples $(H, \xi, b)_1^n$ with an appropriated probability distribution of the components, and the components $H_i$, $\xi_i$ and $b_i$ must satisfy $\xi_i G_{b_i} = H_i$ for all $i=1,\ldots,n$. In order to get this, V realizes the following steps:

1. V chooses randomly, with uniform distribution, a bit $d$.
2. V chooses, with uniform distribution, a random isomorphism $\xi:G_d \to H$.
3. V calculates $h(H)$ and uses its bits to choose $U_c$ and obtaining $|\varphi\rangle=U_c|\theta\rangle$.
4. V teleports $|\varphi\rangle$ to Eve and obtain the bit $b$. She compares $b$ to $d$. If they are equal, what happens with probability ½, V sends $(H, \xi, d)$ to Eve. If they are not equal, V takes a collision $H'$, $h(H')=h(H)$, calculates $\xi'$ such that $\xi' G_{\bar{d}} = H'$ and she sends $(H', \xi', \bar{d})$ to Eve.

As can be seen from step 4, once more in order to have success in cheating Eve, V must know $n$ pairs of collisions $(H, H')$ of the Hash function $h$ and $n$ pairs of isomorphism $(\xi, \xi')$ such that $\xi:G_0 \to H$ and $\xi':G_1 \to H'$. At this moment it is important to stress some points. Firstly, if $b$ is different of $d$ in the protocol just described, then V can decide to send a different pair of classical bits to Eve. However, if this is the case, Eve will obtain a teleported quantum state different from the one that she sent to V, and she will note this in the measurement with probability ¾. Hence, it is better to Eve to send the correct $b$ information and use a collision. In this case, since V does not know what state Eve has sent to her, a collision between two bit sequences $C_1$ and $C_2$ occurs only if $S_1=S_2$, where

$S_{1,2} = \{(|0\rangle, U_{c_{1,2}}|0\rangle), (|1\rangle, U_{c_{1,2}}|1\rangle), (|+\rangle, U_{c_{1,2}}|+\rangle), (|-\rangle, U_{c_{1,2}}|-\rangle)\}$.

This new quantum attack has as advantage over the protocol described in Section 1, the fact that the number of Bell state needed is only $n$, instead of $n^2$. The disadvantage of this new protocol is the present difficulty to implement a Bell state measurement used in the teleportation protocol. At last, once more if a permutation is used instead of a Hash function then attacks using collisions will not be possible.

### III. A QUANTUM ATTACK FOR TRANFERENCE OF PROOF OF ZERO-KNOWLEDGE SYSTEMS WITHOUT USING ENTANGLED STATES

Now let us move to different problem. Let us suppose that Charlie has a secret, a sequence of $n$ bits, $S_C$. He divides the secret in two parts, $S_A$ and $S_B$, such that $S_A \oplus S_B = S_C$. The new secret $S_A$ is delivered to Alice while $S_B$ is delivered to Bob. Furthermore, Charlie sends to Alice a sequence of $m$ qubits $S_q = |\psi_1\rangle \otimes |\psi_2\rangle \otimes \ldots \otimes |\psi_m\rangle$ where $|\psi_i\rangle$ ($i=1,2,\ldots,m$) belongs to the set $\{|0\rangle, |1\rangle, |+\rangle, |-\rangle\}$. At last, the $i$-th bits of $S_A$ and $S_B$ represent, respectively, the binary value that the qubit $|\psi_i\rangle$ represents and the binary value that its basis represents. The states $|0\rangle$ and $|+\rangle$ represent a bit '0' while $|1\rangle$ and $|-\rangle$ represent a bit '1'. The rectangular basis represents the bit '0' and the diagonal basis represents a bit '1'. Now, the task of Alice is to identify Bob, to send to him $S_A$ and to prove to Charlie that she has in fact done it. On the other hand, the Bob's task is to prove to Alice that he is really Bob, to receive $S_A$ from Alice and do not permit Alice discover $S_B$. Hence, as can be realized, the problem just described requires transference of proof of a zero-knowledge system. The solution is the protocol described below:

1. Alice, Bob and Charlie agree in a function $f$, a $k$-bit permutation.
2. Alice splits her $m$ qubit sequence in $k$ smaller sequences having $m/k$ qubits, $s_1, s_2, \ldots, s_k$.
3. Alice sends a qubit sequence $s_i$ to Bob and she asks him to calculate and to send her back the $n/k$ bits of $S_A$ that $s_i$ represents, $S_{Ai}^{(k)}$, the parity of $f\left(S_{Ai}^{(k)} \oplus S_{Bi}^{(k)}\right)$, $b_i$, and the qubit sequence $s_i$.
4. The step 3 is repeated for all $s_i$ ($i=1,\ldots,k$) sequences and, at the end, if Bob provided all bits of $S_A$ correctly, Alice believes that he is really Bob.
5. In order to prove that Alice has found the correct Bob, Alice sends back to Charlie the $k$ bits of parity $(b_1,\ldots,b_k)$ received from Bob and the $m$ qubit sequence.
6. Since Charlie knows $S_A$, $S_B$ and $S_q$ he can always check if the bit $(b_1,\ldots,b_k)$ is correct and if the qubit sequence that he received from Alice is the same that he sent to her at the beginning of the protocol. Hence, he will always know if Alice in fact has found the corrected Bob.

The true Bob knows the bases that he has to use in order to measure correctly the bits that the states sent by Alice represent. For a polarization encoded qubit, Bob can measure the qubit value choosing correctly the polarization beam splitter and using single-photon detectors. In this case, the photons are destroyed and Bob has to produce new ones in the same quantum states that he received from Alice. Another possibility is Bob to use a quantum non-demolition measurement. In this case, since he knows what basis has to be used, he can identify the polarization (bit value) without collapsing the wavefunction and destroying the photon.

## IV. Conclusions

Firstly, we proposed a quantum attack for transference of proofs of a zero-knowledge system using teleportation. The main advantage is the fact that only $n$ pairs of Bell states are necessary and the disadvantage is the hardness to implement Bell state measurement with present technology. Following, it was proposed a quantum attack for transference of proof of a zero-knowledge system without using Bell states. This is possible because, in contrast with the other attacks described in this paper, Eve interacts with verifier and prover before their interaction. At last, this last attack can be implemented with today technology.